\documentclass[a4paper]{article}
\usepackage{url, graphicx, color, varioref, amsfonts, longtable, float}

\newcommand\ASTART{\bigskip\noindent\begin{minipage}[c]{0.5\linewidth}}

\newcommand\AENDSKIP{\end{minipage}\bigskip}
\newcommand\AEND{\end{minipage}}

\newcommand{\qed}{\nobreak \ifvmode \relax \else
      \ifdim\lastskip<1.5em \hskip-\lastskip
      \hskip1.5em plus0em minus0.5em \fi \nobreak
      \vrule height0.75em width0.5em depth0.25em\fi}

\title{A Fast Quantum Algorithm for the Affine Boolean Function Identification}

\author{Ahmed Younes\footnote {ayounes2@yahoo.com or ayounes@alexu.edu.eg}\\
Department of Mathematics and Computer Science, \\Faculty of Science, Alexandria University, \\Alexandria, Egypt 
\and School of Computer Science, University of Birmingham, \\Birmingham, B15 2TT, United Kingdom}

\begin{document}
\maketitle

\begin{abstract}
Bernstein-Vazirani algorithm (the one-query algorithm) can identify a completely specified 
linear Boolean function using a single query to the oracle with certainty. 
The first aim of the paper is to show that if the provided Boolean function is affine, 
then one more query to the oracle (the two-query algorithm) is 
required to identify the affinity of the function with certainty. 
The second aim of the paper is to show that if the provided Boolean function is 
incompletely defined, then the one-query and the two-query algorithms 
can be used as bounded-error quantum polynomial algorithms to identify 
certain classes of incompletely defined linear and affine Boolean functions 
respectively with probability of success at least $2/3$.

\noindent
PACS{03.67.Ac,03.67.Lx,03.65.Yz} 


\end{abstract}

\maketitle

\section{Introduction}

The oracle identification problem is to determine which oracle we have from a set of possible 
Boolean oracles \cite{Ambainis2002}. Another related problem is the oracle property 
testing problem, where the task is to determine if a given oracle has a certain property. 
The complexity of both problems is usually measured by the minimum number of times it is 
required to query the oracle to accomplish that task.   

The case when the oracle represents a linear (affine) Boolean function 
has a special importance. The linearity (and nonlinearity) of Boolean functions is used in 
cryptography, data encryption, error control codes, etc. \cite{Ref11,Ref14,Ref24}. 
Such analysis requires the fully defined form of a Boolean function. 
The incompletely defined Boolean functions, where the correct output for 
certain input vectors are missing, have many applications in 
synthesis and optimization of circuit design. When the function is 
provided in an incompletely defined form, it is important to design efficient methods to 
construct the completely specified form for the incompletely defined Boolean functions.

Classically, it is hard to predict if a given large incompletely defined Boolean function 
can be realized as affine. Many techniques such as spectral techniques have been used 
to analyze Boolean functions in many areas such as classification, testing and evaluation of 
logic complexity, checking if a partially defined Boolean function can be realized in an affine form 
\cite{Ref4,Ref19,Refpaper}.

Designing quantum algorithms for completely specified Boolean functions have gained much attention 
in the literature. The oracle identification problem is solved for linear Boolean functions 
by Bernstein-Vazirani algorithm using a single query to the oracle \cite{Bern1993}. 
The oracle property testing problem to check if a given Boolean oracle is either constant or balanced 
is solved by Deutsch-Jozsa algorithm using a single query to the oracle \cite{Deu1992}. 
In \cite{Dom2010}, quantum algorithms based on the 
Bernstein-Vazirani algorithm \cite{Bern1993} for finding the variables used in a Boolean 
function are presented. 
In \cite{Hill2011}, a quantum algorithm is shown to 
test the linearity of Boolean function using Bernstein-Vazirani algorithm and 
an amplitude amplification technique. In \cite{Kau2013}, an enhanced algorithm of \cite{Hill2011} 
is proposed using Deutsch-Jozsa \cite{Deu1992} and Grover's algorithm \cite{Grov1997}. 
In \cite{Hong2014}, a quantum algorithm for determining the linear structures of a Boolean
function using Bernstein-Vazirani's algorithm and the Simon's algorithm \cite{Sim1997} is presented.

Bernstein-Vazirani's algorithm (the one-query algorithm) is known to identify the linear Boolean function with 
certainty using a single query to the oracle.
It has been noticed by \cite{CLEVE1998,Mont2012} that if the provided function is for an affine Boolean function, 
then Bernstein-Vazirani's algorithm will be blind to the shift experienced by the affinity of the function 
where the affinity of the function will be relegated to an unobservable global phase. 
To overcome this drawback, an independent query to the oracle using $f(0^n)$ would identify the affinity of the 
function. This two independent queries scenario is sufficient to indentify the affine function only if it is 
provided in a completely specified form. 

The aim of the paper is to propose a single algorithm that requires 
two queries to the oracle similar to the above scenario. The proposed algorithm can identify the affine 
Boolean function if it is provided in either a completely specified form 
or in an incompletely defined form. The first aim of this paper is to show that if the given oracle represents 
an affine Boolean function, then one more query to the oracle (the two-query algorithm) is sufficient to identify 
the affine Boolean function with certainty. The second aim is to show that the one-query algorithm 
can identify the linear Boolean function even if the function is provided as an incompletely 
defined function for certain class of functions with probability of success 
at least $2/3$, and the two-query algorithm can identify 
certain class of incompletely defined affine Boolean functions with probability of success 
at least $2/3$.

The paper is organized as follows: Section 2 reviews the basic definitions. 
Section 3 proposes the one-query algorithm and the two-query algorithm for 
the completely specified Boolean functions and the incompletely defined Boolean functions. 
Section 4 gives a discussion about the performance of the one-query algorithm 
and the two-query algorithm respectively. The paper ends up with a conclusion in Section 5.

\section{Basics}

A Boolean function $f$ with $n$ inputs is a mapping $f:X^n \to X$, where $X=\{0,1\}$, 
i.e. the domain of $f$ is the set of $2^n$ binary vectors $\left(0,0,\ldots,0\right)$, 
$\left(0,0,\ldots,1\right)$,$\ldots$, $\left(1,1,\ldots,1\right)$, and $f$ maps each of these vectors to 
the constant 0 or 1. If the domain $B$ of Boolean 
function $f$ is $X^n$ then $f$ is called completely specified Boolean function. 
If $B \subset X^n$, i.e. some input vectors of the function $f$ belong to the set 
$X^n \backslash B$, then the function is called incompletely defined Boolean function. 

Given an incompletely defined version $g$ of a completely specified Boolean function $f$. 
The input vectors that have a value 0 for $f$ are called $OFF_{f}$ cubes, 
the input vectors that have a value 1 for $f$ are called $ON_{f}$ cubes. 
Let $n_0$ and $n_1$ denote the number of input vectors 
in the sets $OFF_{f}$ and $ON_{f}$ respectively then $n_0+n_1=2^n$. 
The input vectors that have a value 0 for $g$ are called $OFF_{g}$ cubes, 
the input vectors that have a value 1 for $g$ are called $ON_{g}$ cubes, the input vectors with no value assigned 
for $g$ are called don't cares or $DC_{g}$ cubes, 
the input vectors with no value assigned for $g$ and have a value 1 for $f$ 
are called $DC1_g$ cubes, and the input vectors with no value assigned for $g$ and have a value 0 for $f$ 
are called $DC0_g$ cubes. Let $n_0^{'}$ and $n_1^{'}$ denote the number of input vectors in the sets $OFF_g$ and $ON_g$ respectively. 
Let $d$ denotes the number of input vectors in the set $DC_g$, i.e. $card$(DC)=$d$ \cite{Refpaper}. 
Let $d_0$ and $d_1$ denote the number of input vectors in the sets $DC0_g$ and $DC1_g$ respectively, 
then $n_0^{'}+n_1^{'}+d_0+d_1=2^n$, 
$d=d_0+d_1 = 2^n - (n_0^{'}+n_1^{'})$, $n_0=n_0^{'}+d_0$, and $n_1=n_1^{'}+d_1$. If $d_0=0$, 
then $n_0 = n_0^{'}$ and if $d_1=0$ then $n_1 = n_1^{'}$.  The undefined values 
of the Boolean function will be denoted by $'-'$. So, an $n$-input incompletely 
defined Boolean function is a mapping $g:X^n  \to X \cup \{ -\}$ \cite{Refpaper}.

An affine Boolean function with $n$ inputs is a Boolean function that can be represented as follows,

\begin{equation}
\begin{array}{l}
f_A\left( {x_0 ,x_1 , \cdots ,x_{n - 1} } \right) = c_0 x_0  \oplus c_1 x_1  \oplus  \ldots  \oplus c_{n - 1} x_{n - 1}  \oplus c_n ,
\end{array}
\end{equation}

\noindent
where $x_i ,c_i  \in X$, $i = 0,1, \ldots ,n$ and 
$\oplus$ denotes bitwise exclusive-or. The affine Boolean function is fully identified 
if the coefficients $c_i$ are known.

If the coefficient $c_n$ is strictly equal to 0 then the function is called a linear Boolean function 
and it can be represented as follows,

\begin{equation}
f_L\left( {x_0 ,x_1 , \cdots ,x_{n - 1} } \right) = c_0 x_0  \oplus c_1 x_1  \oplus  \ldots  \oplus c_{n - 1} x_{n - 1},
\end{equation}

\noindent
where $x_j ,c_j  \in X$, $j = 0,1, \ldots ,{n-1}$. 
The linear Boolean function is fully identified if the coefficients $c_j$ are known, 
this will be denoted as the bit string $C$, where $C=<c_0 c_1 \ldots c_{n-1}>$.

There are $2^{n+1}$ possible $f_A$ functions 
while there are $2^{n}$ possible $f_L$ functions. 
Both types of functions could be balanced, i.e. truth table contains an equal number of 0's and 1's, 
and both types of functions could be constant in a different way, 
for example, if $c_j=0$ for $0 \le j \le n - 1$, then $f_L=0$ while $f_A=$ 0 or 1 depends on 
the value of $c_n$. The function $f_A$ is constant ($f_A=0$) if $c_i=0$ for $0 \le i \le n$. 
If at least one $c_j \ne 0$, then both $f_L$ and $f_A$ are balanced, i.e. $n_0 = n_1 = N/2$, 
where $N=2^n$. If $g_L$ and $g_A$ represents incompletely defined versions of $f_L$ and $f_A$ 
respectively, then $0 \le d_0,d_1 \le N/2$, $d_0=N/2-n_0^{'}$, and $d_1=N/2-n_1^{'}$.

\begin{center}
\begin{figure*}[htbp]
\begin{center}
\setlength{\unitlength}{3947sp}%
\begingroup\makeatletter\ifx\SetFigFont\undefined
\def\x#1#2#3#4#5#6#7\relax{\def\x{#1#2#3#4#5#6}}%
\expandafter\x\fmtname xxxxxx\relax \def\y{splain}%
\ifx\x\y   
\gdef\SetFigFont#1#2#3{%
  \ifnum #1<17\tiny\else \ifnum #1<20\small\else
  \ifnum #1<24\normalsize\else \ifnum #1<29\large\else
  \ifnum #1<34\Large\else \ifnum #1<41\LARGE\else
     \huge\fi\fi\fi\fi\fi\fi
  \csname #3\endcsname}%
\else
\gdef\SetFigFont#1#2#3{\begingroup
  \count@#1\relax \ifnum 25<\count@\count@25\fi
  \def\x{\endgroup\@setsize\SetFigFont{#2pt}}%
  \expandafter\x
    \csname \romannumeral\the\count@ pt\expandafter\endcsname
    \csname @\romannumeral\the\count@ pt\endcsname
  \csname #3\endcsname}%
\fi
\fi\endgroup
\begin{picture}(4907,967)(672,-824)
\thinlines
\put(4016,-111){\oval(236, 200)[tr]}
\put(4016,-111){\oval(236, 200)[tl]}
\put(4030,-582){\oval(236, 200)[tr]}
\put(4030,-582){\oval(236, 200)[tl]}
\put(1178, 14){\line(-5,-6){235}}
\put(947,-104){\line( 1, 0){329}}
\put(1279, 11){\line( 0,-1){235}}
\put(1279,-224){\line( 1, 0){376}}
\put(1655,-224){\line( 0, 1){235}}
\put(1655, 11){\line(-1, 0){376}}
\put(1656,-107){\line( 1, 0){141}}
\put(1799,128){\line( 0,-1){940}}
\put(1799,-812){\line( 1, 0){611}}
\put(2410,-812){\line( 0, 1){940}}
\put(2410,128){\line(-1, 0){611}}
\put(1650,-585){\line( 1, 0){141}}
\put(1271,-465){\line( 0,-1){235}}
\put(1271,-700){\line( 1, 0){376}}
\put(1647,-700){\line( 0, 1){235}}
\put(1647,-465){\line(-1, 0){376}}
\put(941,-579){\line( 1, 0){329}}
\put(2562, 16){\line( 0,-1){235}}
\put(2562,-219){\line( 1, 0){376}}
\put(2938,-219){\line( 0, 1){235}}
\put(2938, 16){\line(-1, 0){376}}
\put(3080,125){\line( 0,-1){940}}
\put(3080,-815){\line( 1, 0){611}}
\put(3691,-815){\line( 0, 1){940}}
\put(3691,125){\line(-1, 0){611}}
\put(2407,-585){\line( 1, 0){141}}
\put(2413,-104){\line( 1, 0){141}}
\put(2556,-472){\line( 0,-1){235}}
\put(2556,-707){\line( 1, 0){376}}
\put(2932,-707){\line( 0, 1){235}}
\put(2932,-472){\line(-1, 0){376}}
\put(2936,-579){\line( 1, 0){141}}
\put(3693,-105){\line( 1, 0){141}}
\put(3834, 11){\line( 0,-1){235}}
\put(3834,-224){\line( 1, 0){376}}
\put(4210,-224){\line( 0, 1){235}}
\put(4210, 11){\line(-1, 0){376}}
\put(3693,-574){\line( 1, 0){141}}
\put(3834,-465){\line( 0,-1){235}}
\put(3834,-700){\line( 1, 0){376}}
\put(4210,-700){\line( 0, 1){235}}
\put(4210,-465){\line(-1, 0){376}}
\put(4214,-105){\line( 1, 0){141}}
\put(4214,-577){\line( 1, 0){141}}
\put(3912,-158){\vector( 2, 1){282}}
\put(3919,-640){\vector( 2, 1){282}}
\put(2941,-108){\line( 1, 0){141}}
\put(1086, 81){$n$}
\put(1292,-189){$H^{\otimes n}$}
\put(2580,-189){$H^{\otimes n}$}
\put(2640,-660){$H$}
\put(1352,-660){$H$}
\put(3300,-361){$U_f$}
\put(1998,-361){$U_f$}

\put(689,-189){$\left| 0 \right\rangle$}
\put(689,-660){$\left| 1 \right\rangle$}
\put(4411,-189){$\left| c_0,c_1,\ldots,c_{n-1} \right\rangle$}
\put(4411,-660){$\left| c_n \right\rangle$}

\end{picture}%
\end{center}
\caption{A quantum circuit for the proposed two-query algorithm.}
\label{algv2}
\end{figure*}
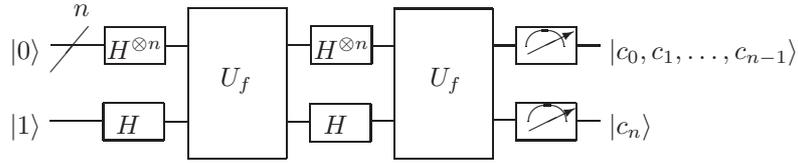
\end{center}

In the literature, a Boolean function is considered as an oracle that marks certain states in a
superposition. There are two ways used to mark the states, 
one way is to conditionally apply certain phase shifts on the
marked states \cite{Grov1997} by using an oracle $V_f$ that works as follows: 
$V_f \left| x \right\rangle  = \left( { - 1} \right)^{f(x)} \left| x \right\rangle$. 
The other way is to use an oracle $U_f$ to entangle the required
states with certain state of the extra qubit workspace \cite{Younes03b} as follows:  
$U_f \left| {x,0} \right\rangle  = \left| {x,f(x)} \right\rangle$, where the state of 
the extra qubit workspace is required for further operations. 
The oracle $U_f$ is used by initializing the $n+1$ qubits quantum register to 
the state $\left| 0 \right\rangle ^{ \otimes n+1}$, then 
apply the operator $H^{\otimes n}\otimes I$ to the register, where $I$ is the $2 \times 2$ 
identity matrix. The oracle $U_f$ can perform as $V_f$ by initializing the $n+1$ qubits 
quantum register to the state $\left| 0 \right\rangle ^{ \otimes n}  \otimes \left| 1 \right\rangle$, then 
apply the operator $H^{\otimes n+1}$ to the register and ignore the extra qubit workspace afterward, where $H$ is the Hadamard gate defined as follows,

\begin{equation}
H = \frac{1}{{\sqrt 2 }}\left[ {\begin{array}{*{20}c}
   1 & 1  \\
   1 & { - 1}  \\
\end{array}} \right].
\end{equation}

Applying the $H$ gate on a qubit in state $\left\vert
0\right\rangle $ or $\left\vert 1\right\rangle $ will produce a qubit in a
perfect superposition. In general, the effect of applying the $H$ gate on an $n$-qubits 
quantum register is known as Walsh-Hadamard transform and can be represented as follows,

\begin{equation}
H^{ \otimes n}\left| x \right\rangle  = \frac{1}{{\sqrt {2^n } }}\sum\limits_{y = 0}^{2^n  - 1} {\left( { - 1} \right)^{x . y} \left| y \right\rangle }, 
\end{equation}

\noindent where $x.y=x_0.y_0 \oplus x_1.y_1 \oplus \ldots \oplus x_{n-1}.y_{n-1}$, and $x_j.y_j$ 
is the bitwise-and between $x_j$ and $y_j$.

\section{The Proposed Algorithm}\label{Sec3}

\subsection{Completely Specified Boolean Function}

Given a quantum register of $n+1$ qubits in state $\left| 0 \right\rangle ^{ \otimes n}\otimes \left| 1 \right\rangle $ 
and an oracle $U_f$ that represents an $n$ inputs completely specified affine Boolean function $f$, 
then the operations of the proposed algorithm $A_1$ (shown in fig. \ref{algv2}) 
can be written as follows,

\begin{equation}
A_1 = U_f H^{n + 1} U_f H^{n + 1}.
\end{equation}

\subsubsection*{Tracing the Algorithm}
The operations of the proposed algorithm can be understood as follows where the first three steps 
are straight forward from Bernstein-Vazirani algorithm, 

\begin{itemize}

\item[1.] Prepare a quantum 
register of $n+1$ qubits, the first $n$ qubits in state $\left| 0 \right\rangle$ and an extra qubit is state 
$\left| 1 \right\rangle$ as follows,
\begin{equation}
\left| {\Psi _0 } \right\rangle  = \left| 0 \right\rangle ^{ \otimes n}  \otimes \left| 1 \right\rangle .
\end{equation}

\item[2.] Apply $H^{ \otimes n+1}$,
\begin{equation}
\begin{array}{l}
 \left| {\Psi _1 } \right\rangle  = \left( {H^{ \otimes {n+1}}  } \right)\left| {\Psi _0 } \right\rangle  \\ 
 \,\,\,\,\,\,\,\,\,\,\,\,\, = \frac{1}{{\sqrt {2^n } }}\sum\limits_{x = 0}^{2^n  - 1} {\left| x \right\rangle }  \otimes \left( {\frac{{\left| 0 \right\rangle  - \left| 1 \right\rangle }}{{\sqrt 2 }}} \right) . \\ 
 \end{array}
\end{equation}

\item[3.] Apply $U_f$ on the $n+1$ qubits,
\begin{equation}
\begin{array}{l}
 \left| {\Psi _2 } \right\rangle  = U_f \left| {\Psi _1 } \right\rangle  \\ 
 \,\,\,\,\,\,\,\,\,\,\,\,\,  = \frac{1}{{\sqrt {2^n } }}\sum\limits_{x = 0}^{2^n  - 1} {\left( { - 1} \right)^{f(x)} \left| x \right\rangle }  \otimes \left( {\frac{{\left| 0 \right\rangle  - \left| 1 \right\rangle }}{{\sqrt 2 }}} \right) . \\ 
 \end{array}
\end{equation}

\item[4.] Apply $H^{ \otimes n+1}$,
\begin{equation}
\begin{array}{l}
 \left| {\Psi _3 } \right\rangle  = \left( {H^{ \otimes {n+1}}  } \right)\left| {\Psi _2 } \right\rangle  \\ 
 \,\,\,\,\,\,\,\,\,\,\,\,\, = \frac{1}{{2^n }}\sum\limits_{x = 0}^{2^n  - 1} {\sum\limits_{z = 0}^{2^n  - 1} {\left( { - 1} \right)^{f(x) + x.z} \left| z \right\rangle } }  \otimes \left| 1 \right\rangle,\\ 
 \end{array}
\end{equation}

\noindent
and since the vectors of the linear Boolean functions (ignoring $c_n$)
form an orthonormal basis, i.e. the following identity holds,
\begin{equation}
 \sum\limits_{x = 0}^{2^n  - 1} (-1)^{x \cdot z} = 2^{n} \delta_{z,0},
\end{equation}
\noindent where $x$ and $z$ are $n$-bit strings, then, $\left| {\Psi _3 } \right\rangle$ can be written as follows \cite{Hill2011},

\begin{equation}
\left| {\Psi _3 } \right\rangle  = \left( { - 1} \right)^{c_n } \left| {c_0 c_1  \ldots c_{n - 1} } \right\rangle  \otimes \left| 1 \right\rangle. 
\end{equation}

It is important to notice that Bernstein-Vazirani algorithm is not sensitive to the 
affinity of the oracle, i.e. the value of $c_n$, where the affinity appears as a global 
phase shift of $\left( { - 1} \right)^{c_n }$ which will not be detected when the quantum 
register is measured. So, one more query to the oracle is required to find the value of 
$c_n$.

\item[5.] To find the value of $c_n$, apply $U_f$ on the $n+1$ qubits \cite{Younes03b},
\begin{equation}
\begin{array}{l}
 \left| {\Psi _4 } \right\rangle  = U_f \left| {\Psi _3 } \right\rangle  \\ 
 \,\,\,\,\,\,\,\,\,\,\,\,\, = \left( { - 1} \right)^{c_n } \left| {c_0 c_1  \ldots c_{n - 1} } \right\rangle  \otimes \left| {1 \oplus c_n  \oplus p_c  } \right\rangle,  \\ 
 \end{array}
\end{equation}
\noindent where $p_c  = c_0  \oplus c_1  \oplus  \ldots  \oplus c_{n - 1} $.

\item[6.] Measure the first $n$ qubits to get the bit string $\left| {c_0 c_1  \ldots c_{n - 1} } \right\rangle$.
\item[7.] Measure the extra qubit to read the value of $c_n$ as $\left| {1 \oplus c_n  \oplus  p_c  } \right\rangle$ 
such that if the number of 1's in the bit string $\left| {c_0 c_1  \ldots c_{n - 1} } \right\rangle$ is even then 
the measured value in the extra qubit is $\left|{1 \oplus c_n }\right\rangle$, i.e. the negation of ${c_n }$, 
and if the number of 1's in the bit string $\left| {c_0 c_1  \ldots c_{n - 1} } \right\rangle$ is odd then
the measured value in the extra qubit is $\left|{c_n }\right\rangle$.

\end{itemize}

\subsection{Incompletely Defined Boolean Function}

Given an $n$ inputs incompletely defined affine Boolean function $g$ as follows,

\begin{equation}
g(x)  = \left\{ {\begin{array}{*{20}l}
   {0} & {\,\,if\,\,x \in OFF_g,}\\
   {1} & {\,\,if\,\,x \in ON_g,}\\
   {2} & {\,\,if\,\,x \in DC_g,}\\
\end{array}} \right.
\label{g_x}
\end{equation}
\noindent where $g(x)=2$ if $x \in DC_g$ represents a third choice for the don't cares. 
To find the completely specified version of $g$, $g(x)=2$ should be replaced with either $g(x)=0$ or $g(x)=1$,  
the correct replacement is not known in advance. 
Quantum parallelism can be exploited to examine both replacements simultaneously. 
This can be done by encoding the third choice, i.e. $g(x)=2$, in a quantum version $U_g$ of the oracle as 
${\textstyle{1 \over {\sqrt 2 }}}\left( {\left| 0 \right\rangle  + \left| 1 \right\rangle } \right)$. 
This can be achieved by assuming that the don't care input vector $x$ is in the set $ON_g$, 
and then replace the $NOT$ gate with the Hadamard gate in the controlled gate representation for the 
minterms equivalent to the don't care input vector $x$ \cite{Younes03b}. 

The proposed algorithm to find the completely specified version of $g$ is as follows: prepare 
a quantum register of $n+1$ qubits in state $\left| 0 \right\rangle ^{ \otimes n}\otimes \left| 1 \right\rangle$ 
and the quantum oracle $U_g$ that represents the $n$ inputs incompletely 
defined affine Boolean function $g$ defined as follows,

\begin{equation}
U_g \left| x \right\rangle  \otimes \left| t \right\rangle  = \left\{ {\begin{array}{*{20}l}
   {\left| x \right\rangle  \otimes \left| t\oplus g(x) \right\rangle } & {if\,\,x \notin DC_g,}  \\
   {\left| x \right\rangle  \otimes H \left| t \right\rangle } & {if\,\,x \in DC_g,}  \\
\end{array}} \right.
\label{oracle1}
\end{equation}

\noindent 
where the don't cares for $g$ are encoded as $\left( {\frac{{\left| 0 \right\rangle  + \left| 1 \right\rangle }}{{\sqrt 2 }}} \right)$, 
then the operations of the proposed algorithm $A_2$ can be written as follows,

\begin{equation}
A_2 = U_g H^{n + 1} U_g H^{n + 1}.
\end{equation}

\subsubsection*{Tracing the Algorithm}
The operations of the proposed algorithm can be understood as follows,

\begin{itemize}

\item[1.] Prepare a quantum register of $n+1$ qubits, the first $n$ qubits in state $\left| 0 \right\rangle$ 
and an extra qubit is state 
$\left| 1 \right\rangle$ as follows,
\begin{equation}
\left| {\Psi _0 } \right\rangle  = \left| 0 \right\rangle ^{ \otimes n}  \otimes \left| 1 \right\rangle .
\end{equation}

\item[2.] Apply $H^{ \otimes n+1}$,
\begin{equation}
\begin{array}{l}
 \left| {\Psi _1 } \right\rangle  = \left( {H^{ \otimes {n+1}}  } \right)\left| {\Psi _0 } \right\rangle  \\ 
 \,\,\,\,\,\,\,\,\,\,\,\,\, = \frac{1}{{\sqrt {2^n } }}\sum\limits_{x = 0}^{2^n  - 1} {\left| x \right\rangle }  \otimes \left( {\frac{{\left| 0 \right\rangle  - \left| 1 \right\rangle }}{{\sqrt 2 }}} \right). \\ 
 \end{array}
\end{equation}

\item[3.] Apply $U_g$,

\begin{equation}
\begin{array}{l}
\left| {\psi _2 } \right\rangle  = U_g \left| {\psi _1 } \right\rangle\\
 \,\,\,\,\,\,\,\,\,\,\,\,\,  = \frac{1}{{\sqrt {2^n } }}\sum\limits_{\scriptstyle x = 0, \hfill \atop 
  \scriptstyle x \notin DC \hfill}^{2^n  - 1} {\left( { - 1} \right)^{g(x)} \left| x \right\rangle }  \otimes \left( {\frac{{\left| 0 \right\rangle  - \left| 1 \right\rangle }}{{\sqrt 2 }}} \right) 
+ \frac{1}{{\sqrt {2^n } }}\sum\limits_{\scriptstyle x = 0, \hfill \atop 
  \scriptstyle x \in DC \hfill}^{2^n  - 1} {\left| x \right\rangle }  \otimes \left| 1 \right\rangle.\\ 
\end{array}
\end{equation}

\item[4.] Apply $H^{ \otimes n+1}$. To simplify calculations, first apply $I^{ \otimes n} \otimes H$,

\begin{equation}
\begin{array}{l}
 \left| {\psi _3 } \right\rangle  = \left( {I^{ \otimes n}  \otimes H} \right)\left| {\psi _2 } \right\rangle  \\ 
  \,\,\,\,\,\,\,\,\,\,\,\,\,= \frac{1}{{\sqrt {2^n } }}\sum\limits_{\scriptstyle x = 0, \hfill \atop 
  \scriptstyle x \notin DC \hfill}^{2^n  - 1} {\left( { - 1} \right)^{g(x)} \left| x \right\rangle }  \otimes \left| 1 \right\rangle 
+ \frac{1}{{\sqrt {2^n } }}\sum\limits_{\scriptstyle x = 0, \hfill \atop 
  \scriptstyle x \in DC \hfill}^{2^n  - 1} {\left| x \right\rangle }  \otimes \left( {\frac{{\left| 0 \right\rangle  - \left| 1 \right\rangle }}{{\sqrt 2 }}} \right) \\ 
  \,\,\,\,\,\,\,\,\,\,\,\,\,= \frac{1}{{\sqrt {2^n } }}\sum\limits_{\scriptstyle x = 0, \hfill \atop 
  \scriptstyle x \notin DC \hfill}^{2^n  - 1} {\left( { - 1} \right)^{g(x)} \left| x \right\rangle }  \otimes \left| 1 \right\rangle 
   + \frac{1}{{\sqrt {2^{n + 1} } }}\sum\limits_{\scriptstyle x = 0, \hfill \atop 
  \scriptstyle x \in DC \hfill}^{2^n  - 1} {\left| x \right\rangle }  \otimes \left| 0 \right\rangle  
 - \frac{1}{{\sqrt {2^{n + 1} } }}\sum\limits_{\scriptstyle x = 0, \hfill \atop 
  \scriptstyle x \in DC \hfill}^{2^n  - 1} {\left| x \right\rangle }  \otimes \left| 1 \right\rangle  \\ 
 \end{array}
\end{equation}
 
\noindent
then apply $H^{ \otimes n} \otimes I$,

\begin{equation}
\label{linpart}
\begin{array}{l}
 \left| {\psi _4 } \right\rangle  = \left( {H^{ \otimes n}  \otimes I} \right)\left| {\psi _3 } \right\rangle  \\ 
    \,\,\,\,\,\,\,\,\,\,\,\,\,= \frac{1}{{2^n }}\sum\limits_{\scriptstyle x = 0, \hfill \atop 
  \scriptstyle x \notin DC \hfill}^{2^n  - 1} {\sum\limits_{z = 0}^{2^n  - 1} {\left( { - 1} \right)^{g(x) + x.z} } \left| z \right\rangle }  \otimes \left| 1 \right\rangle  
   + \frac{1}{{2^n }}\sum\limits_{\scriptstyle x = 0, \hfill \atop 
  \scriptstyle x \in DC \hfill}^{2^n  - 1} {\sum\limits_{z = 0}^{2^n  - 1} {\left( { - 1} \right)^{x.z} } \left| z \right\rangle }  \otimes \left( {\frac{{\left| 0 \right\rangle  - \left| 1 \right\rangle }}{{\sqrt 2 }}} \right) \\ 
    \,\,\,\,\,\,\,\,\,\,\,\,\,= \sum\limits_{\scriptstyle z = 0, \hfill \atop 
  \scriptstyle z \ne C \hfill}^{2^n  - 1} {\alpha _z \left| z \right\rangle }  \otimes \left| 0 \right\rangle  + \sum\limits_{\scriptstyle z = 0, \hfill \atop 
  \scriptstyle z \ne C \hfill}^{2^n  - 1} {\beta _z \left| z \right\rangle }  \otimes \left| 1 \right\rangle  
+ \gamma _0 \left| C \right\rangle  \otimes \left| 0 \right\rangle  + \gamma _1 \left| C \right\rangle  \otimes \left| 1 \right\rangle,  \\ 
 \end{array}
\end{equation}

\noindent
where, 
\begin{equation}
\begin{array}{l}
\gamma _0  = \frac{{\left( { - 1} \right)^{c_n } }}{{\sqrt{2} \,\, 2^n  }}\left( {d_0  - d_1 } \right), 
\gamma _1  = \left( { - 1} \right)^{c_n } \left( 1 - {\frac{{\left( \sqrt 2 d + d_0  - d_1\right) }}{{\sqrt{2}\,\, 2^n  }}} \right).
\end{array}
\end{equation}
\end{itemize}

If it is sufficient to find the linear part of the function, then the probability of success to get 
the linear part correctly is $P_L  = \gamma _0 ^2  + \gamma _1 ^2$. If it is required to find the affinity 
of the function, i.e. the value of $c_n$, then apply $U_g$ one more time.  
To simplify the calculations and since we are interested in the bit string $C$, 
so the subsystem $\left| {\psi _C } \right\rangle$ of interest is as follows,

\begin{equation}
\left| {\psi _C } \right\rangle  = \gamma _0 \left| C \right\rangle  \otimes \left| 0 \right\rangle  + \gamma _1 \left| C \right\rangle  \otimes \left| 1 \right\rangle.
\end{equation}

We have to consider if $C \in DC_g$ or not. If $C \notin DC_g$ then applying $U_g$ gives,
\begin{equation}
\begin{array}{l}
\left| {\psi _{C(final)} } \right\rangle  = U_g \left| {\psi _C } \right\rangle\\
\,\,\,\,\,\,\,\,\,\,\,\,\,\,\,\,\,\,\,\,\,\,\,\,\,\,\,\,\,\, =\gamma _0 \left| C \right\rangle  \otimes \left| { c_n   \oplus p_c } \right\rangle 
+ \gamma _1 \left| C \right\rangle  \otimes \left| {1 \oplus c_n   \oplus p_c } \right\rangle,
 \end{array}
 \label{eqncnotug}
 \end{equation}

and if $C \in DC_g$ then applying $U_g$ gives,
  
\begin{equation}
\begin{array}{l}
 \left| {\psi _{C(temp)} } \right\rangle  = U_g \left| {\psi _C } \right\rangle  \\ 
  \,\,\,\,\,\,\,\,\,\,\,\,\,\,\,\,\,\,\,\,\,\,\,\,\,\,\,\,\,= \gamma _0 \left| C \right\rangle  \otimes \left( {\frac{{\left| 0 \right\rangle + \left( { - 1} \right)^{c_n } \left| 1 \right\rangle }}{{\sqrt 2 }}} \right)
  + \gamma _1 \left| C \right\rangle  \otimes \left( {\frac{{\left| 0 \right\rangle  - \left( { - 1} \right)^{c_n } \left| 1 \right\rangle }}{{\sqrt 2 }}} \right) \\ 
  \,\,\,\,\,\,\,\,\,\,\,\,\,\,\,\,\,\,\,\,\,\,\,\,\,\,\,\,\,  = \frac{1}{{\sqrt 2 }}\left( {\gamma _0  + \gamma _1 } \right)\left| C \right\rangle  \otimes \left| {0 } \right\rangle  
   + \frac{{\left( { - 1} \right)^{c_n } }}{{\sqrt 2 }}\left( {\gamma _0  - \gamma _1 } \right)\left| C \right\rangle  \otimes \left| {1} \right\rangle. \\ 
 \end{array}
 \label{eqncinug}
\end{equation}

To increase the probability of success of finding $c_n$ in $\left| {\psi _{C(temp)} } \right\rangle$ 
when $C \in DC_g$, then apply $(I^{ \otimes n} \otimes H)$,

\begin{equation}
\begin{array}{l}
 (I^{ \otimes n}  \otimes H)\left| {\psi _{C(temp)} } \right\rangle  = \frac{1}{{\sqrt 2 }}\left( {\gamma _0  + \gamma _1 } \right)\left| C \right\rangle  \otimes \left( {\frac{{\left| 0 \right\rangle  + \left| 1 \right\rangle }}{{\sqrt 2 }}} \right) + \frac{{\left( { - 1} \right)^{c_n } }}{{\sqrt 2 }}\left( {\gamma _0  - \gamma _1 } \right)\left| C \right\rangle  \otimes \left( {\frac{{\left| 0 \right\rangle  - \left| 1 \right\rangle }}{{\sqrt 2 }}} \right) \\ 
  \,\,\,\,\,\,\,\,\,\,\,\,\,\,\,\,\,\,\,\,\,\,\,\,\,\,\,\,\,= \frac{1}{2}\left( {\gamma _0  + \gamma _1  + \left( { - 1} \right)^{c_n } \left( {\gamma _0  - \gamma _1 } \right)} \right)\left| C \right\rangle  \otimes \left| 0 \right\rangle  + \frac{1}{2}\left( {\gamma _0  + \gamma _1  - \left( { - 1} \right)^{c_n } \left( {\gamma _0  - \gamma _1 } \right)} \right)\left| C \right\rangle  \otimes \left| 1 \right\rangle  \\ 
  \,\,\,\,\,\,\,\,\,\,\,\,\,\,\,\,\,\,\,\,\,\,\,\,\,\,\,\,\,=\gamma _0 \left| C \right\rangle  \otimes \left| {c_n   \oplus p_c } \right\rangle 
+ \gamma _1 \left| C \right\rangle  \otimes \left| {1 \oplus  c_n   \oplus p_c } \right\rangle,
 \end{array}
\end{equation}

\noindent 
which is equal to $\left| {\psi _{C(final)} } \right\rangle$ for $C \notin DC_g$ shown in eq. (\ref{eqncnotug}). 
So, $\left| {\psi _{C(final)} } \right\rangle$ can be taken as the final state of the system in analyzing the 
probability of success of finding $c_n$. We can assume that the probability of success 
to get the coefficients of the affine Boolean function is $P_A  = \gamma _1 ^2$, since $\gamma _0 \le \gamma _1$, where 
the value of $c_n$ can be read as $\left| {1 \oplus c_n  \oplus  p_c  } \right\rangle$ similar to 
the case of completely specified functions.

\section{Discussion}

\begin{figure*}[htbp]
\centerline{\includegraphics{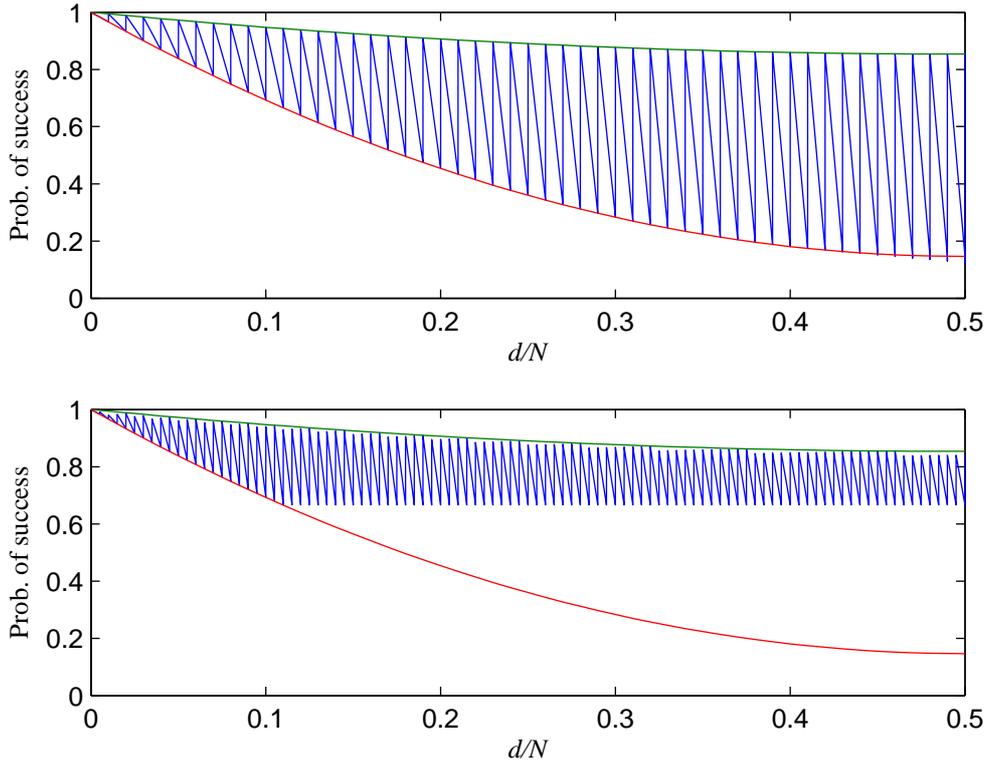}}
\caption{ a. (up) The probability of success for the spectrum of incompletely defined linear functions, where the upper bound for $d=d_1$ and the lower bound for $d=d_0$, 
b. (down) The probability of success for the class of incompletely defined linear functions according to the condition shown in eq. (\ref{condlinear}).} 
\label{lpart}
\end{figure*}

The probability of success to get the coefficients of the incompletely defined linear 
Boolean function is,

\begin{equation}
\begin{array}{l}
 P_L = \gamma _0 ^2  + \gamma _1 ^2  \\ 
\,\,\,\,\,\,\,\,\,\,  = \left( {\frac{{d_0  - d_1 }}{{\sqrt{2} 2^n  }}} \right)^2  + \left( 1 - {\frac{{\left( \sqrt 2 d + d_0  - d_1\right) }}{{\sqrt{2} 2^n  }}} \right)^2,  \\ 
 \end{array}
\end{equation}

\noindent such that $d=d_0+d_1$, $0 \le d < N/2$, and $0 \le d_0,d_1 < N/2$. 
Let $D=d/N$, $D_0=d_0/N$, and $D_1=d_1/N$, such that, 
$0 \le D < 1/2$, and $0 \le D_0,D_1 < 1/2$. So, $P_L$ can be written as follows,

\begin{equation}
\label{problinear}
\begin{array}{l}
P_L  = \left( {1 - \left( {1 + \frac{1}{{\sqrt 2 }}} \right)D + \sqrt 2 D_1 } \right)^2  + \left( {\frac{1}{{\sqrt 2 }}D - \sqrt 2 D_1 } \right)^2. 
\end{array}
\end{equation}

The probability of success for the spectrum of incompletely defined linear Boolean functions 
depends on the number of don't cares. If the set $DC_g$ contains only 
members from $DC1_g$, i.e. $D=D_1$, then $0.85 \le P_L \le 1$, and if 
the set $DC_g$ contains only members from $DC0_g$, i.e. $D=D_0$, then $0.15 \le P_L \le 1$ 
as shown in fig. \ref{lpart}(a).

The class of incompletely defined linear Boolean functions for which the one-query algorithm 
can succeed with probability at least 2/3 as shown in fig. \ref{lpart}(b), 
i.e. $P_L\ge 2/3$ must satisfy the following condition,

\begin{equation}
\label{condlinear}
\begin{array}{l}
D_1  \ge \frac{{\sqrt {K_1 ^2  + 4K_2 }  - K_1 }}{4} \ge 0,
D_0  \le D - \frac{{\sqrt {K_1 ^2  + 4K_2 }  - K_1 }}{4} \ge 0,
\end{array}
\end{equation}

\noindent where $ K_1  = \sqrt 2  - (2 + \sqrt 2 )D$, and $K_2  = (2 + \sqrt 2 )D(1 - D) - 1/3$.

For $d \ge N/2$, the oracle $U_g$ might be equivalent to more than one completely specified linear Boolean function, 
since the Hamming distance between the truth table of any two completely specified linear Boolean functions 
is equal to $N/2$. For example, if $d=N/2$, then $U_g$ is equivalent to two completely specified linear Boolean 
functions $f_1(x)$ and $f_2(x)$, then $\left| {\psi _4 } \right\rangle$ in eq. (\ref{linpart}) 
can be re-written as follows (ignoring the affinity of $f_1(x)$ and $f_2(x)$),

\begin{equation}
\left| {\psi _4 } \right\rangle  = \frac{1}{2}\left( {\left| {C_1 } \right\rangle  + \left| {C_2 } \right\rangle } \right) \otimes \left| 1 \right\rangle  + \frac{1}{2}\left( {\left| 0 \right\rangle ^{ \otimes n}  - \left| {C_1  \oplus C_2 } \right\rangle } \right) \otimes \left( {\frac{{\left| 0 \right\rangle  - \left| 1 \right\rangle }}{{\sqrt 2 }}} \right),
\end{equation}

\noindent where $C_1$ and $C_2$ are the bit strings for $f_1(x)$ and $f_2(x)$ respectively. The probability of success 
to get $C_1$ or $C_2$ is $1/4$ with a total probability of success of $1/2$ which is 
outside the class of incompletely defined linear Boolean functions for which the one-query algorithm 
can succeed with probability at least 2/3.

\begin{figure*}[htbp]
\centerline{\includegraphics{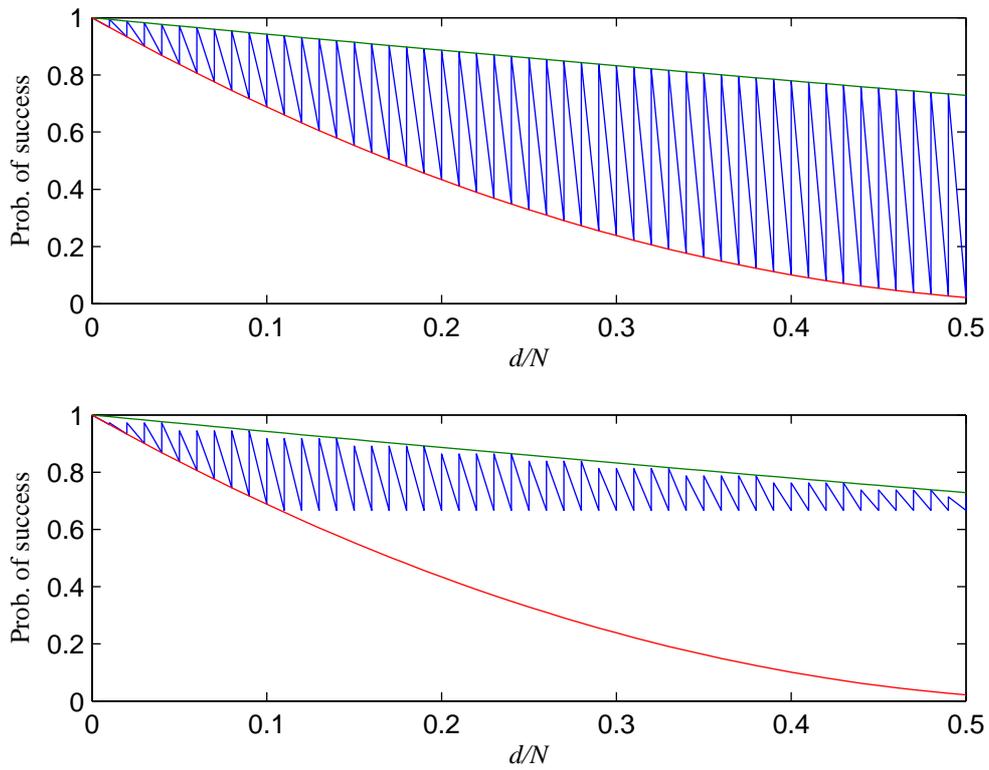}}
\caption{ a. (up) The probability of success for the spectrum of incompletely defined affine functions, where the upper bound for $d=d_1$ and the lower bound for $d=d_0$, 
b. (down) The probability of success for the class of incompletely defined affine functions according to the condition shown in eq. (\ref{condaffine}).} 
\label{apart}
\end{figure*}

The probability of success to get the coefficients of the incompletely defined 
affine Boolean function is,

\begin{equation}
\begin{array}{l}
\label{probaffine}
 P_A  = \gamma _1^2  = \left( {1 - \frac{{\left( {\sqrt 2 d + d_0  - d_1 } \right)}}{{2^n \sqrt 2 }}} \right)^2  \\ 
 \,\,\,\,\,\,\,\,\,\, = \left( {1 - \left( {1 + \frac{1}{{\sqrt 2 }}} \right)D + \sqrt 2 D_1 } \right)^2  \\ 
 \end{array}
\end{equation}

The probability of success for the spectrum of incompletely defined affine Boolean functions 
depends on the number of don't cares. If the set $DC_g$ contains only 
members from $DC1_g$, i.e. $D=D_1$, then $0.72 \le P_A \le 1$, and if 
the set $DC_g$ contains only members from $DC0_g$, i.e. $D=D_0$, then $0.02 \le P_A \le 1$ 
as shown in fig. \ref{apart}(a).
  
The class of incompletely defined affine Boolean functions for which the two-query 
algorithm can succeed with probability at least 2/3 as shown in fig. \ref{apart}(b), 
i.e. $P_A\ge 2/3$ must satisfy the following condition,

\begin{equation}
\label{condaffine}
\begin{array}{l}
D_1  \ge \frac{1}{{\sqrt 3 }} - \frac{1}{{\sqrt 2 }} + \left( {\frac{{1 + \sqrt 2 }}{2}} \right)D \ge 0,
D_0  \le \frac{1}{{\sqrt 2 }} - \frac{1}{{\sqrt 3 }} + \left( {\frac{{1 - \sqrt 2 }}{2}} \right)D \ge 0.
\end{array}
\end{equation}

It can be seen that the probability of success shown in eqs. (\ref{problinear}) and (\ref{probaffine}) 
favor the don't cares that belong to the set $DC1_g$ over the don't cares 
that belong to the set $DC0_g$, i.e. the probability of success is higher if the don't cares 
in the provided incompletely defined Boolean function are supposed to be the value 1 in the 
corresponding completely specified Boolean function. The reason is that the oracle $U_g$ used, 
as shown in eq. (\ref{oracle1}), 
is mapping the don't care vectors to the state ${\textstyle{1 \over {\sqrt 2 }}}\left( {\left| 0 \right\rangle  + \left| 1 \right\rangle } \right)$. 
If the oracle maps the don't care vectors to the state 
${\textstyle{1 \over {\sqrt 2 }}}\left( {\left| 0 \right\rangle  - \left| 1 \right\rangle } \right)$ 
instead, then the algorithms will favor the don't cares that belong to the set $DC0_g$. 
This can be done by encoding the third choice, $g(x)=2$, in a quantum version $U_g^{'}$ of the oracle as 
${\textstyle{1 \over {\sqrt 2 }}}\left( {\left| 0 \right\rangle  - \left| 1 \right\rangle } \right)$. 
This can be achieved by assuming that the don't care input vector $x$ is in the set $OFF_g$, 
and then replace the $NOT$ gate with the Hadamard gate in the controlled gate representation for the 
minterm equivalent to the don't care input vector $x$ \cite{Younes03b}. 
If the values of $n_0^{'}$ and $n_1^{'}$ are known, then the values of $d_0$ and $d_1$ can be calculated respectively, 
i.e. $d_0=N/2-n_0^{'}$, and $d_1=N/2-n_1^{'}$. If $d_0 < d_1$ then the oracle shown 
in eq. (\ref{oracle1}) is used in the algorithms, otherwise the following oracle is used instead,   

\begin{equation}
U_g^{'} \left| x \right\rangle  \otimes \left| t \right\rangle  = \left\{ {\begin{array}{*{20}l}
   {\left| x \right\rangle  \otimes \left| t\oplus g(x) \right\rangle } & {if\,\,x \notin DC_g,}  \\
   {\left| x \right\rangle  \otimes H(NOT \left| t \right\rangle) } & {if\,\,x \in DC_g.}  \\
\end{array}} \right.
\label{oracle2}
\end{equation}

The ability to choose the correct oracle will double the number of the incompletely defined Boolean 
functions in the class of functions for which the algorithm can succeed with probability at least $2/3$. 
If the values of $n_0^{'}$ and $n_1^{'}$ are not known, then the algorithms may run 
constant number of times using each of the oracles $U_g$ and $U_g^{'}$ in turn, 
then the winner with more votes in the majority vote from the two runs 
is taken as the correct output string.


\section{Conclusion}

Bernstein-Vazirani algorithm (the one-query algorithm) is known to identify a completely specified linear 
Boolean function using a single query to the oracle with certainty. 
It has been shown that Bernstein-Vazirani algorithm is not sensitive to the 
affinity of the oracle. So, one more query to the oracle is required after 
Bernstein-Vazirani algorithm (the two-query algorithm) to be able to identify 
a completely specified affine Boolean function with certainty.
 
The one-query algorithm and the two-query algorithm are also able to identify classes of incompletely 
defined Boolean functions with probability at least $2/3$. The probability of success depends 
on the number of don't cares and on the choice to encode the don't care in the oracle  
as ${\textstyle{1 \over {\sqrt 2 }}}\left( {\left| 0 \right\rangle  + \left| 1 \right\rangle } \right)$ 
or ${\textstyle{1 \over {\sqrt 2 }}}\left( {\left| 0 \right\rangle  - \left| 1 \right\rangle } \right)$.

\end{document}